\documentclass{vgtc}                          % final (conference style)
\ifpdf%                                % if we use pdflatex
  \pdfoutput=1\relax                   % create PDFs from pdfLaTeX
  \usepackage{graphicx}                % allow us to embed graphics files
  \DeclareGraphicsExtensions{.pdf,.png,.jpg,.jpeg} % for pdflatex we expect .pdf, .png, or .jpg files
\else%                                 % else we use pure latex
  \usepackage{graphicx}                % allow us to embed graphics files
  \DeclareGraphicsExtensions{.eps}     % for pure latex we expect eps files
\fi%

\graphicspath{{figures/}{pictures/}{images/}{./}} % where to search for the images

\usepackage{microtype}                 % use micro-typography (slightly more compact, better to read)
\PassOptionsToPackage{warn}{textcomp}  % to address font issues with \textrightarrow
\usepackage{textcomp}                  % use better special symbols
\usepackage{mathptmx}                  % use matching math font
\usepackage{times}                     % we use Times as the main font
\usepackage{cite}                      % needed to automatically sort the references
\usepackage{tabu}                      % only used for the table example
\usepackage{booktabs}                  % only used for the table example
\usepackage{eurosym}
\usepackage{enumitem}

\usepackage{amsfonts}
\usepackage{amsmath}
\usepackage{tcolorbox}

\newcommand*\samethanks[1][\value{footnote}]{\footnotemark[#1]}
\onlineid{1}

\vgtccategory{Research}

\vgtcinsertpkg

\title{Dynamic Field of View Reduction Related to Subjective Sickness Measures in an HMD-based Data Analysis Task \looseness=0}

\author{Daniel Zielasko\thanks{e-mail: \{zielasko, freitag, weyers, kuhlen\}@vr.rwth-aachen.de}
\and Alexander Mei{\ss}ner\thanks{e-mail: alexander.meissner@rwth-aachen.de}
\and Sebastian Freitag\samethanks[1]
\and Benjamin Weyers\samethanks[1]
\and Torsten W. Kuhlen\samethanks[1]}
\affiliation{\scriptsize Visual Computing Institute, RWTH Aachen University, Germany\\JARA-HPC, Aachen, Germany }
\abstract
{
Various factors influence the degree of cybersickness a user can suffer in an immersive virtual environment, some of which can be controlled without adapting the virtual environment itself.
When using HMDs, one example is the size of the field of view.
However, the degree to which factors like this can be manipulated without affecting the user negatively in other ways is limited. 
Another prominent characteristic of cybersickness is that it affects individuals very differently.
Therefore, to account for both the possible disruptive nature of alleviating factors and the high interpersonal variance, a promising approach may be to intervene only in cases where users experience discomfort symptoms, and only as much as necessary.
Thus, we conducted a first experiment, where the field of view was decreased when people feel uncomfortable, to evaluate the possible positive impact on sickness and negative influence on presence.
While we found no significant evidence for any of these possible effects, interesting further results and observations were made.
} % end of abstract

\CCScatlist{
    \CCScat{Human-centered concepts}{Human computer interaction (HCI)}{Interaction paradigms}{Virtual reality}{};
    \CCScat{Human-centered concepts}{Human computer interaction (HCI)}{Visualization}{Empirical studies in visualization}{}
}

\begin{document}


\firstsection{Introduction\label{sec::intro}}

\maketitle
Today, virtual reality (VR) increasingly integrates into our lives, be it for entertainment or in professional contexts \cite{zielasko2017, zielasko2017buenos, Sousa2017}.
This may result in people being exposed for long periods to immersive virtual environments (IVEs) in an uncontrolled way.
While longer exposure times have not yet been sufficiently studied (one of the rare examples is \cite{steinicke2014}) it is already well known that the symptoms of people that suffer from cybersickness \cite{laviola2000} increase over time.
Broadly accepted as the main reason for cybersickness are sensory conflicts \cite{reason1975}, which can result from being immersed into the IVE while simultaneously being exposed to potentially deviating real world physical behaviors.
Typical symptoms are nausea, dizziness and headache, which can last several hours or even days \cite{Stanney2003} and thus can have a significant impact on the fun and/or the productivity, next to the pure impact on the well-being of the user. 
The group of people that are negatively affected is substantial and actually may increase with the time of usage \cite{Stanney2003}. 
Nevertheless, some of the factors that influence the degree of sickness a user suffers from can be controlled.
When using HMDs, one example is the size of the field of view (FoV) \cite{dizio1997, lin2002, fernandes2016}.
However, the degree in which it can be manipulated, such that they do not have other negative effects, is limited too.
While these negative effects, such as a reduced presence \cite{seay2001, cummings2016, lin2002}, might be acceptable or even unrecognizable for someone who otherwise would get sick, it is restricting for someone who does not exhibit any symptoms.
Even though it is possible to predict the vulnerability of a person from some individual factors a priori, like gender, fitness, prior experience etc. \cite{Kennedy1995}, the variance is big and it might additionally change even for a single person.
Therefore, Zielasko et al. \cite{zielasko2017buenos} introduced the concept to utilize user profiles to manage cybersickness countermeasures.
For every user, a profile is created, initially based on a set of demographic questions and experiences.
Those variables together with the users' reactions to the experience then influence possible sickness countermeasures, such as a decrease of the FoV.
In the best case, the initial predictions lose more and more influence while the system learns how to deal with this individual user.
But the described system today is a vision and in practice this vision still offers a bunch of challenges that have to be addressed.
%Thus, as the contribution of this work we conducted a study to test the general feasibility of this vision.
Thus, as the main contribution of this work we want to address one of these challenges, namely if it is possible to reduce cybersickness by radically reducing the FoV depending on the individual well-being without its common drawbacks.
Therefore, we designed and implemented a prototypical and streamlined version of the proposed system.
In a seated data analysis task (the scenario matches the one defined by the term \textit{deskVR} \cite{zielasko2017}), the FoV of the study group's participants adapted with respect to their repeatedly self-reported well-being, while the control group experienced an unmodified FoV.

In the following section \ref{sec::rw}, related work regarding the avoidance and reduction of cybersickness and the possibilities to measure cybersickness is brought together.
Then, the implementation of the FoV reduction, which was used in the user study, is presented in section \ref{sec::FoV}. 
Afterwards, the user study, the used apparatus, task, procedure, the participants and finally our hypotheses are presented in section~\ref{sec::study}.
Its results are given in section~\ref{sec::results} and discussed in section~\ref{sec::discussion}.
Finally, the work is concluded in section \ref{sec::conclusion}.

\section{Sickness, Countermeasures and Measures\label{sec::rw}}
Cybersickness is influenced by many factors, some closer related to sensory conflict theory \cite{reason1975} than others.
These factors can be distinguished into two groups: properties of the individual (that cannot be changed, but considered), and system properties. 
An example of an individual factor is the user's experience with VR that is negatively correlated with cybersickness \cite{Bailenson2006, Freitag2016, cobb1999}.
Here the user could be slightly introduced into the IVE by step-wise increased exposure times \cite{Stanney2003}, when circumstances permit.
The factors given by the system are potentially controllable and further split into three classes.
The first class consists of mandatory ones, like framerate, latency and jitter \cite{Meehan2003, Wilson2016}.
They are as fundamental and their impact on sickness is as high that every VR setup should first aim for a sufficiently high framerate, a precise tracking and low latency.
Second, there are similar factors like utilizing body cues for travel \cite{zielasko2016}  that has shown to prevent sickness, but as they usually do not come with disadvantages just want to be maximized.
None of the two factors really can be utilized as a dynamic cybersickness countermeasure and worse even the combination of all factors might be not sufficient to totally eliminate sickness.
But there is a last class of factors, those which it might be worth to vary as they have a negative impact as well.
Only the last are possible countermeasures for a profiled system.
One example for this is the FoV provided by the system.
It is correlated to cybersickness and can have positive effects when it is decreased \cite{dizio1997, lin2002}.
Additionally, the FoV can be changed when using an HMD (see section \ref{sec::FoV}).
But different from the other mentioned factors, it is usually the goal to maximize the FoV, which should lead to higher immersion and thus presence \cite{seay2001, cummings2016}, which then again correlates with sickness \cite{witmer1998,lin2002, llorach2014}.
Thus, a general decrease of the FoV might have either positive or negative effects on sickness but in any case reduces immersion.
Therefore, Fernandez and Feiner \cite{fernandes2016} dynamically and subtly decreased the FoV with respect to the movement and rotation of the user such that they should not notice it, leading to reduced negative effects.
The perception threshold is different from person to person, it is for instance known that females tend to have a higher FoV \cite{Kennedy1985}, but this dynamic adaption may be also integrated by default as it potentially serves no drawbacks. 
Nevertheless, the change is by design just subtle and so it effects potentially are subtle, as well.
This raises the question whether the FoV can be decreased much more radically, when the user starts to feel sick without the same drastic negative effects, as the sickness might cover other side-effects or the potential decrease in sickness compensates them.
This would raises one major need.
How to continuously get a reliable insight to the sickness level, respectively well-being, of the user, such that it is possible to react to it?
Research has found some correlations between biophysiological measurements and cybersickness \cite{kim2005, Nalivaiko2015, Dennison2016} that are comparatively easy to obtain, but those are very sparsely examined.
The work of Kim et al. \cite{Kim2008bio} serves as a first example for a system that actually changed the FoV and warned the user to take a break in reaction to changes of biosignals.  
In contrast, questionnaires like the simulator sickness questionnaire (SSQ) \cite{Kennedy1993} are  well investigated and well established.
However, it is not possible to track the current well-being of a user with it and additionally the SSQ is better suited for the comparison of populations than to judge the absolute state of well being of a single individual, which is usually the case for subjective measures like this. 
Thus, in this work we connected the size of the FoV to the state of well-being of the participants by just regularly asking for a rating on a simple scale \cite{fernandes2016, Freitag2017}, knowing that this is only a compromise in continuation as well as in reliability and additionally can disturb the user and such has to be tackled in the future (see section~\ref{sec::discussion}). 

\section{Field of View Reduction\label{sec::FoV}}
\begin{figure}[tb]
	\centering
	\includegraphics[width=\columnwidth]{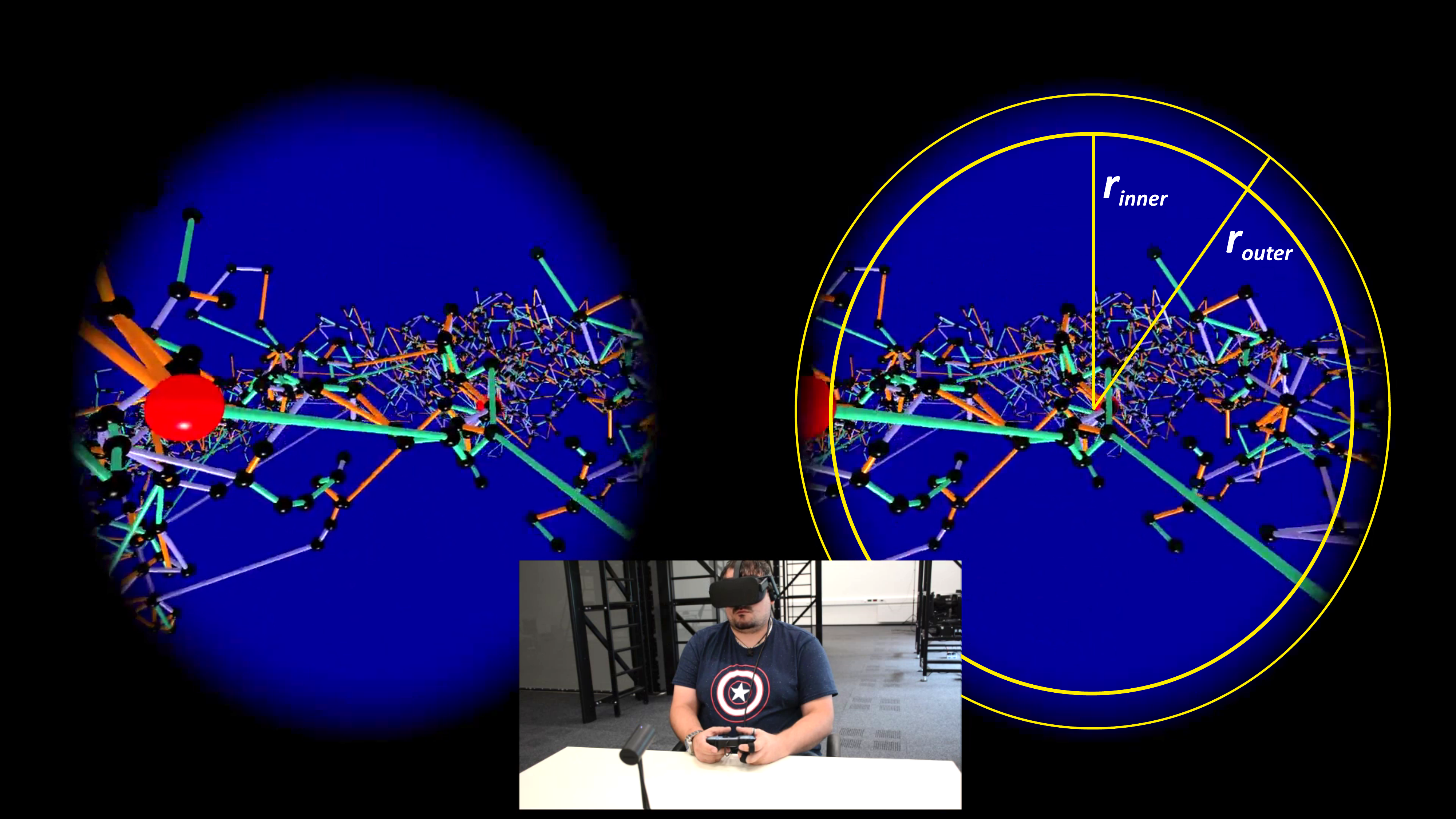}
	\caption{\label{FIG::realDesk} A screenshot of the stereo image sent to the HMD: Depicted  are the graph and the slightly larger and red colored pair of spheres, which the participant has to find the shortest path between. The right half is annotated with indicators for the inner and outer radii of the FoV limitation. Additionally, the screenshot is overlayed with a photo of the experimental setup for the user study on the bottom.}
\end{figure}

In the following, the technical implementation of the FoV reduction is described.
Different from the method used by Fernandez and Feiner \cite{fernandes2016}, we do not use a textured quad as a mask but a black full-screen quad.
The cutout and the blending into it are realized by a GLSL pixel shader that varies the transparency values.
Therefore, the cutout circle, given by the radius $r_{inner}$, and the transition circle surrounding it, given by $r_{outer}$ (see Figure \ref{FIG::realDesk}), are given by uniform variables with all position data given in screen-space coordinates. This gives more control over the appearance of the FoV limitation.
The actual transparency $\alpha$ of a pixel $f$ is given by its distance $d_f$ to the center as:
\begin{equation*}
\alpha (f)=
\begin{cases}
    0,& \text{if } d_f < r_{inner}\\
    1,& \text{if } d_f > r_{outer}\\
    \frac{d_f - r_{inner}}{r_{outer} - r_{inner}},& \text{otherwise}
\end{cases} 
\end{equation*}

\noindent This is mirrored for both eyes.
The variables $r_{inner}$ and $r_{outer}$ are varied to achieve the actual dynamic FoV reduction (see section \ref{sec::study::fov}).
An example of the FoV limitation is illustrated in Figure~\ref{FIG::realDesk}.

\section{User Study\label{sec::study}}
The study tasks were inspired by \cite{zielasko2016}, where the participants had to use different techniques to repeatedly travel through a large graph (5214 vertices and 6913 edges) in node-link representation and find a highlighted pair of vertices to identify the shortest path between (see section \ref{sec::study::task}).
In the study that is described here the travel technique was kept constant and was a standard 4 DoF gamepad control, i.e., roll rotation and vertical translation were not enabled.
However, the available FoV was varied between the participants while in the study group there was an automatic field of view reduction (see section \ref{sec::FoV} and \ref{sec::study::fov}) that was influenced by a personal health state, which was updated through the subject during the experiment (see section \ref{sec::study::task}), in the control group, the health state was also updated by the user but it induced no effects. 

\subsection{Apparatus\label{sec::study::apparatus}}
For all participants, the experiment took place in front of an regular office desk, while they were seated on a tiltable office chair.
The virtual environment was projected by an Oculus Rift Consumer Version 1.
The used gamepad was an Logitech RumblePad\textsuperscript{TM}~2.
Furthermore, all participants were equipped with a chest strap to record the heart rate and two finger electrodes to measure the skin resistance.
These data is gathered as part of a long-term experiment and thus is not evaluated here.

\subsection{Task\label{sec::study::task}}
In the beginning of each task, the participant was instructed by a text in the VE to find the shortest path between two highlighted vertices.
To make the task more difficult, further conditions were often added.
For this purpose, each edge was randomly colored, using a color out of a color blind safe set of three colors (green, violet, and orange), using the same coloring scheme for all participants. 
Two examples for additional conditions are, for instance, \textit{find the shortest path with at least two orange edges} or \textit{find the shortest path without any violet edge}.
Every participant received the same tasks in the same order, with generally increasing task difficulty.
As soon as a participant was convinced to have solved the task she triggered any button on the gamepad to open a dialog.
This dialog contained 10 spheres on a horizontal line each labeled with a value, starting from one to ten.
To answer the task a participant had to look at the corresponding sphere, which changed its color as feedback, and simultaneously trigger any button on the gamepad.
We chose this method over, e.g., a preselected sphere, where the selection is moved to the right or the left with the sticks of the gamepad to reduce possible bias.
After answering, the participant was asked in a second dialog for her well-being (in the following referred to as \textit{health score}), again on a scale from 1 (good) to 10 (bad).
Finally, the next task was specified in a dialog, which had to be confirmed by pressing any button.

\subsection{Field of View Settings\label{sec::study::fov}}
For the study group, the FoV was varied between $110^{\circ}$ (diagonal), which is the maximum FoV of the used HMD, and $44^{\circ}$ at minimum.
Furthermore, every health score was linearly mapped to a specific FoV in this range.
When the FoV had to change this happened with a constant speed of $0.44^{\circ}/s$.
The FoV started at its maximum and did not change for first two minutes of the experiment, i.e. for the whole training.
In order to keep the results clean, the FoV was only adjusted according to the health score, i.e., the travel speed or rotation speed did not have any influence.

\subsection{Procedure\label{sec::study::procedure}}
Every participant of the experiment went to the following procedure.
First, everybody was asked to sign an informed consent form.
Second, the participants were non-anonymously asked, if they already had experienced VR.
Based on their answer and gender they were evenly distributed between the study condition and the control condition, as several studies have shown a strong correlation between those two factors and the sensitivity to cybersickness \cite{Kennedy1995, Freitag2016, Bailenson2006}.
In the following, a written description of the study tasks and procedure was handed out with the request to be carefully read and the notice to ask the experimenter as soon as any question arises.
They were told that the experiment is about spatial interaction in data analysis applications.
Additionally, the participants were asked to give preference to precision over speed in the experiment.   
Afterwards, every participant anonymously filled out a demographic questionnaire and an \textit{a priori} simulator sickness questionnaire (SSQ) \cite{Kennedy1993}.
Subsequent to all the paperwork, which took about 15 minutes, the participants were equipped with the chest strap and electrodes (see section \ref{sec::study::apparatus}) and then seated in front of the desk.
The participants were exposed exactly 15 minutes to the virtual environment, of which the first 2 minutes were reserved for practicing the controls and the procedure.
The time was restricted to keep the cybersickness scores comparable, while the participants themselves were not aware that the study is terminated after 15 minutes.
After the experimental part of the study, everybody was asked to anonymously fill out some concluding questionnaires including Likert-scale items, an \textit{a posteriori} SSQ and a Witmer and Singer presence questionnaire \cite{witmer1998}.
The whole procedure took about 45 minutes per person.
The textual content, both on paper and in the IVE, was presented in both English and German side by side. 

\subsection{Participants\label{sec::study::participants}}
33 participants (8 female and 25 male, mean age 27.3, SD = 4.7) took part in the experiment.
Three of them (2 female, 1 male) canceled the study before its end due to nausea.
All three were part of the control condition, i.e., did not receive any cybersickness countermeasures.
The data of all three was excluded from the analysis.
It was possible to rebalance the conditions during the study, which at the end still led to an even distribution regarding the number of participants, gender and experience with VR.
All participants were unpaid and were only compensated with free candy and soft drinks.

\subsection{Hypotheses\label{sec::study::hypotheses}}
The experiment was mainly designed to, evaluate the general feasibility of sickness counter-measures based on repeatedly measuring the subjective degree of sickness and, second, to find out if a more radical decrease in the FoV based on those measures has a negative impact on factors like presence.
Thus, the FoV of the study condition's participants changed with respect to their subjective well-being.
Therefore, we expect the following effect.
\begin{itemize}[leftmargin = 0.9cm, itemsep = 0.0cm]
\item[\textbf{H1}] The participants in the study group experience less cybersickness then the control group during and after the experiment. 
\end{itemize}
Furthermore, we assumed that the reduction of the FoV will be noticed but simultaneously does not hamper the participants as it reduces their sickness, even though they might not be aware of this link.
In summary, we did not expect a negative effect on presence, as the FoV only changes for participants who feel sick.
\begin{itemize}[leftmargin = 0.9cm, itemsep = 0.0cm]
\item[\textbf{H2}] The degree of presence experienced by the participants in the study group is not different from the control group.
\end{itemize} 
\vspace{-1.em}
\begin{itemize}[leftmargin = 1.5cm, itemsep = 0.0cm]
\item[\textbf{H2.1}] The study group notices the field of view reduction.
\item[\textbf{H2.2}] The study group does not feel disturbed by the field of view reduction.
\end{itemize} 

\section{Results\label{sec::results}}
\begin{figure}[tb]
	\centering
	\includegraphics[width=\columnwidth]{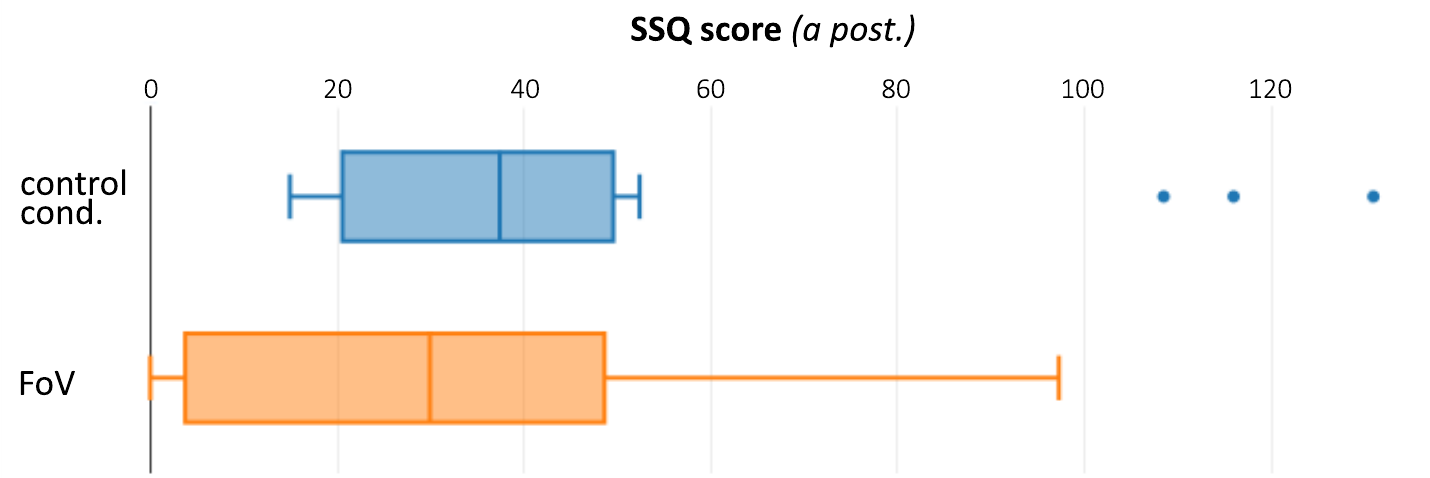}
	\caption{\label{FIG::box} Box plots of the SSQ scores for the study and control condition.}
\end{figure}

We report all results using a significance level of $.05$ and non-significant trends at a level of $.1$. 
We analyzed the results with independent-samples t-tests, using Welch-Satterthwaite adjustments to the degrees of freedom instead where Levene's test indicated that the assumption of homogeneity of variances was violated.
The results of the core parameters are given in Table \ref{tab::main} and the SSQ score is additionally depicted in Figure \ref{FIG::box}.

\subsection{Core Measures\label{sec::results::measures}}
Regarding the core measures, the t-tests revealed no significant difference between the study and the control group for the SSQ scores, the subjective sickness measures, the presence score and the performance parameters, i.e., number of correctly solved tasks and the ratio of errors (see Table \ref{tab::main}).
Additionally, some travel parameters were recorded and evaluated.
Here we found that the study group traveled a significantly longer distance $T(28) = -2.180, p = .038$ and looked around to a higher degree, $T(28) = -2.526, p = .017$.
Last, the amount of virtual rotation, performed with the gamepad, did not differ significantly.
\begin{table}[tb]
\centering
\caption{The results for the core measures for the study group (\textit{FoV}) and the control group, in addition to, the p-values of the comparing t-tests. The $\Delta$SSQ score is the difference between an \textit{a priori} SSQ and the \textit{a posteriori} SSQ. The same holds for the $\Delta$health score.}
\label{tab::main}
\scriptsize
\begin{tabular}{lcrrr}
                       & \textbf{Cond.}	& \textbf{Mean}		& \textbf{Std. Dev.} & \textbf{Sig.}	\\ \hline
SSQ score \textit{(a post.) } 	   &     	& 47.6      & 38.5 & 			\\
				                   & FoV   	& 31.7      & 29.3 & .212\phantom{$^{*}$}		\\ \hline
$\Delta$ SSQ score       		   &     	& 35.7	    & 35.2 & 			\\
				                   & FoV   	& 20.7 	    & 24.9 & .190\phantom{$^{*}$}		\\ \hline
last health score 			   &    	& 4.7       & 2.4  &         	\\
			                       & FoV   	& 3.4       & 2.9  & .197\phantom{$^{*}$} 	    \\ \hline
$\Delta$  health score    	   &     	& 2.9  	    & 1.9  &			\\
			                       & FoV   	& 2.3  	    & 2.8  & .542\phantom{$^{*}$}	    \\ \hline
presence    	   				   &     	& 95.8  	& 17.0 &			\\
			                       & FoV   	& 98.0  	& 28.8 & .813\phantom{$^{*}$}       \\ \hline
\# correct tasks   	   			   &     	& 10.7  	& 4.7  &  			\\
			                       & FoV   	& 11.7  	& 2.7  & .483\phantom{$^{*}$}		\\ \hline
task error \textit{in} \%    	   		   &     	& 22.3  	& 23.4 &  			\\
			                       & FoV   	& 17.0  	& 8.1  & .417\phantom{$^{*}$}		\\ \hline
distance traveled \textit{in} $m$   	   		   &     	& 1,343.4  	& 462.1&  			\\
			                       & FoV   	& 1,772.8  	& 606.8& \textbf{.038}$^{*}$		\\ \hline
head rotation \textit{in} $^{\circ}$		   &     	& 8,259.5  	& 4,082.4&			\\
			                       & FoV   	& 12,720.4  	& 5,488.7& \textbf{.017}$^{*}$		\\ \hline
virtual rotation \textit{in} $^{\circ}$     &     	& 6,583.6  	& 3,789.4& 			\\
			                       & FoV   	& 8,885.6  	& 5,304.6& .182\phantom{$^{*}$}		\\
			                                            
\end{tabular}
\end{table}

\begin{figure}[tb]
	\centering
	\includegraphics[width=\columnwidth]{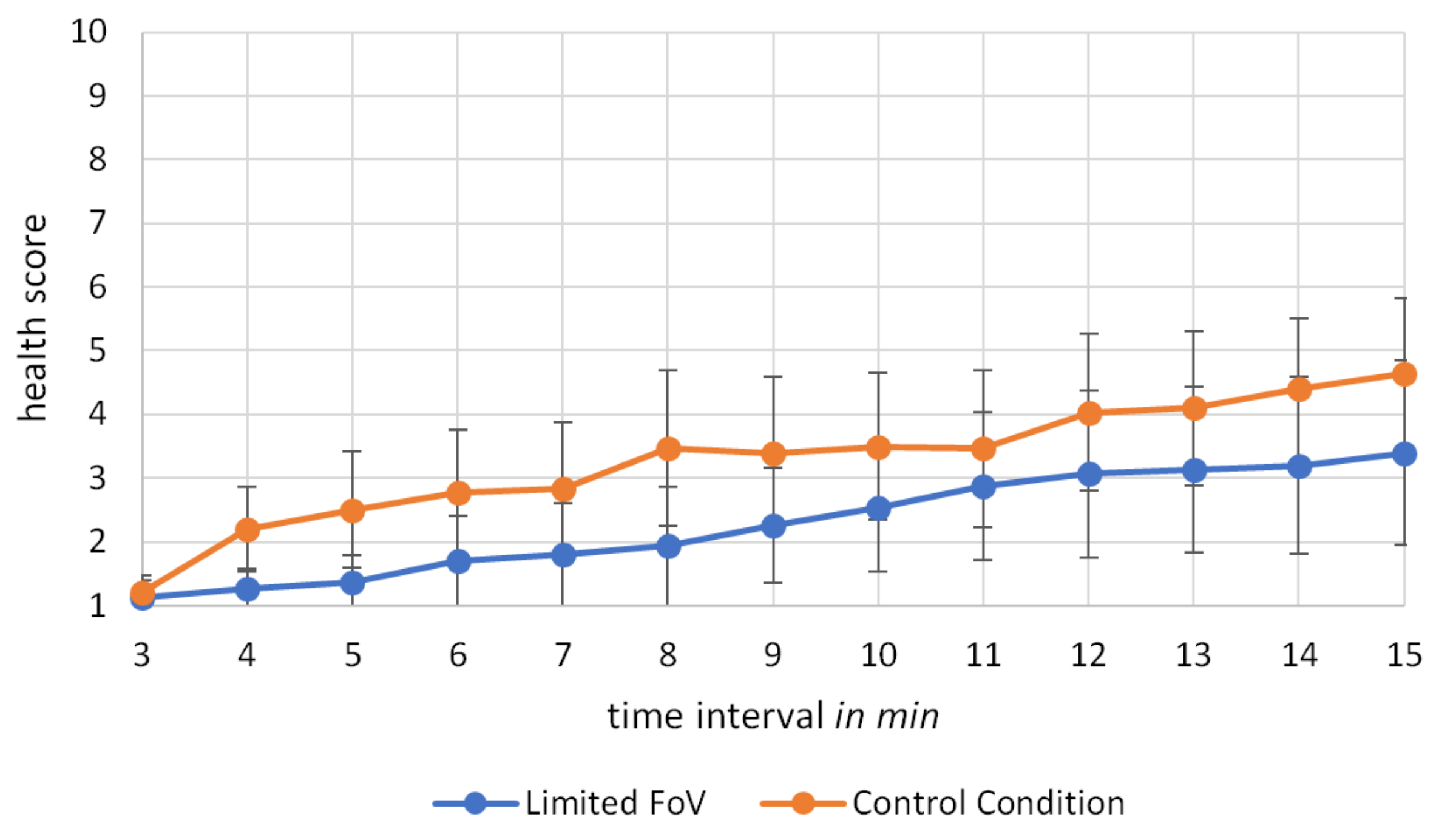}
	\vspace{-2.em}
	\caption{\label{FIG::series} History of the health score (ranging from 1 (good) to 10 (bad)) updated by the participants during the experiments for the study group (\textit{limited FoV}) and the control group. The series is given starting with minute 3, excluding the training. Error bars show the 95\% confidence
	intervals.}
\end{figure}

Furthermore, the continuously gathered subjective health score for every user was binned to a specific minute resulting in series of measurements (see Figure \ref{FIG::series}), repeating the rating whenever there was no update in some minute.
This series was analyzed with a two-way mixed-design ANOVA with a between-subjects factor of dynamic FoV reduction (enabled, disabled). 
However, there was no statistical difference found between the two groups, $F(12, 336) = .846, p = .603)$.

\subsection{Subjective Measures\label{sec::results::quest}}
\begin{figure}[tb]
	\centering
	\includegraphics[width=\columnwidth]{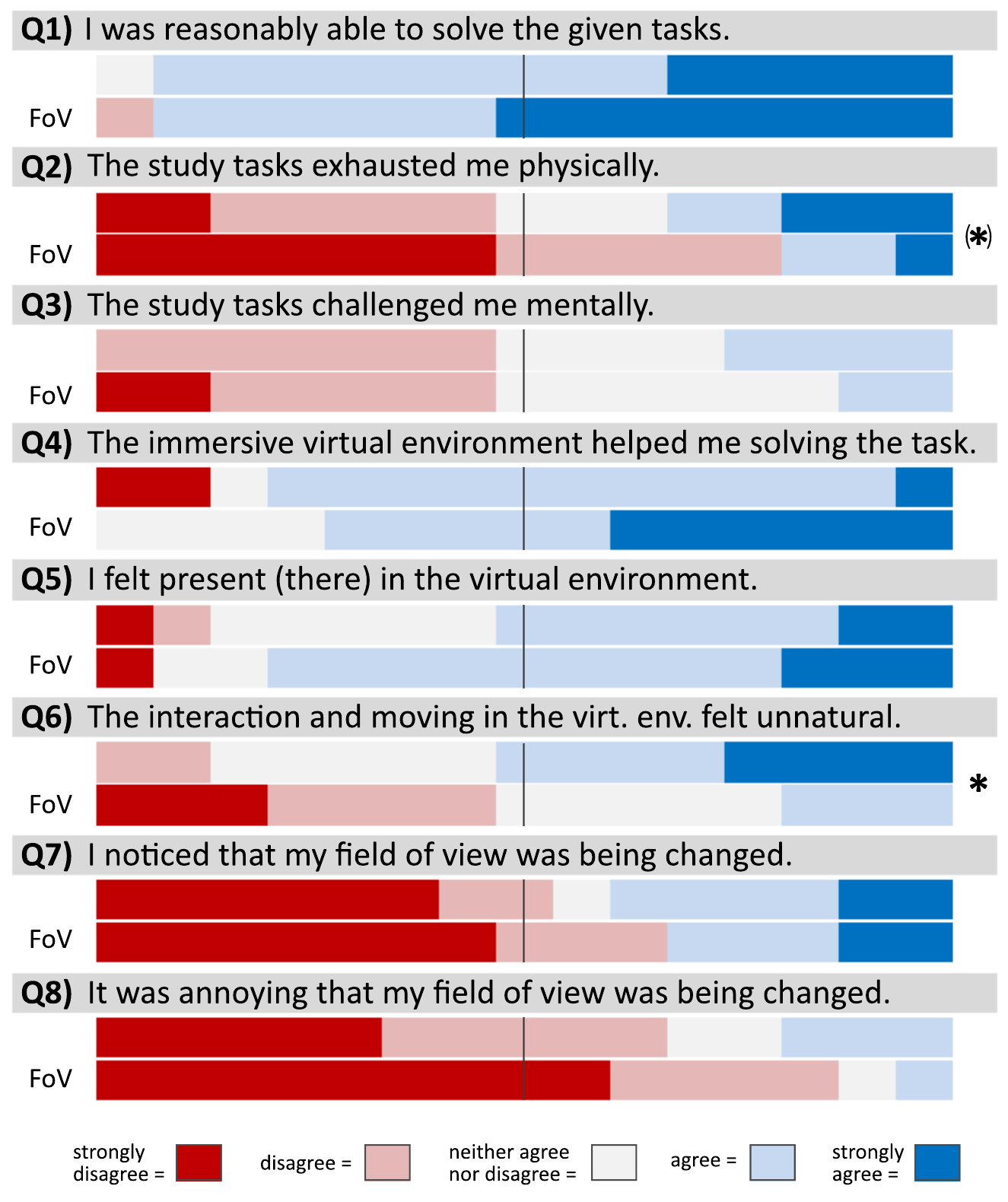}
	\caption{\label{FIG::quest} The answers to the Likert-scale items (1-5) for the study condition (\textit{FoV}) and the control condition. An asterisk marks a statistically significant difference ($p<.05$) and a star in brackets an statistical trend ($p<.1$).}
\end{figure}

The results of the Likert-scale items are given in Figure \ref{FIG::quest}.
All but two items, showed no statistically significant difference in the answers given by the participants.
First, for the control group the interaction in the IVE felt significant more unnatural (\textbf{Q6}), $T(28) = 2.947, p = .006$.
Second, there was a statistical trend indicating that the control group felt more exhausted physically by the tasks (\textbf{Q2}), $T(28) = 1.895, p = .068$.

\subsection{Correlation Analysis\label{sec::results::corr}}
A Pearson product-moment correlation coefficient was computed to assess the relationship between the measures. 
We especially focused on possible correlations between individual factors and sickness, as well as the behavior of well-established measure such as the SSQ to the subjective health questions.
The tests found no correlation between the age, the experience with VR or the subjective vulnerability to motion sickness to any cybersickness measure.
However, there was a strong positive correlation between the subjective vulnerability to cybersickness and all sickness measures, e.g., for the SSQ score, $r(30) = .496, p = .005$ and the last health question, $r(30) = .539, p = .002$.
Finally, there was a strong positive correlation between all sickness measures, e.g., \textit{a posteriori} SSQ score and last health questions, $r(30) = .674, p < .001$.

\section{Discussion\label{sec::discussion}}
First to note is that we were not able to confirm \textbf{H1}, expecting that the study group feels less sick then the control group after the experiments.
None of the sickness measures showed a significant difference.
However, we are confident that the reduction of the FoV had an effect, which is reflected in the fact that the study group traveled a significantly longer distance and looked around significantly more while simultaneously showing no significant difference in the number of tasks correctly solved and errors made.
Another explanation would be that the partially reduced FoV decreased the grade of presence as much that the positive effects on sickness, as discussed in section \ref{sec::rw}, were counterbalanced \cite{seay2001, cummings2016}.
But neither the presence scores \cite{witmer1998} nor the Likert-scale item Q5 show a difference between the conditions and their felt presence, which additionally confirms \textbf{H2}.
Furthermore, the participants of the study group seem to not have noticed the FoV reduction at all, even though it was very drastic (down to about 48$^{\circ}$) for some participants, as Q7 and Q8 suggest.
Thus, \textbf{H2.1} can surprisingly not be confirmed and might suggest that the FoV may be reduced even more drastically, faster or earlier when people feel sick.
This outcome then confirms \textbf{H2.2}, i.e., people were not disturbed by the restricted FoV.
We think that a big factor for the results is that the FoV was reduced with a delay, as the participants first had to finish a task and in the following could report an increased sickness.
Additionally, the FoV started at its maximum and was not reduced in the training.
Even though we had a reason to do so, namely that the participants should be able to freely explore the interfaces, an already increasing discomfort was not counteract during this period. 
In summary, we think that the results are promising, but that the FoV reduction was maybe still to cautious and possible effects might be covered by the participants that did not feel any sickness at all.
A small indication for these effects are suggested by the Likert-scale items Q2 and Q6.
For the latter, the control group rated the interaction significantly less natural.
This might relate to presence, but we are not completely sure how to interpret this result.
For Q2 there was a statistical trend suggesting that the control group felt more physically exhausted by the tasks.
Thus, the experiment should be repeated with more participants, a more drastic FoV reduction regarding amount, speed and sensitivity and a longer exposure time, as the overall degree of sickness was less than expected.

\subsection{Health Score}
The conditions from this experiment are not directly usable in practice.
The participants were repeatedly asked to rate their well-being during the study.
Although we found a strong correlation between the health score and the established sickness measures, the possibility to repeatedly ask the user in a standard application how they feel is usually not given, as it interrupts the user.
Furthermore, we are not sure if the self-perception is always reliable and sufficiently fast.
The reliability may be increased by asking the user only if her well-being changed and if yes, in which direction.
This may take an unnecessary burden from the user and is as sufficient as an absolute rating for the system, we think. 
However, as mentioned before, this might only be a solution for further research but cannot be a final solution, as well as giving the responsibility of updating the score to the user.   
Thus, for real applications, biophysiological measurements might be an answer, but existing research has to be extended \cite{Kim2008bio, kim2005} and sensors have to be less cumbersome.
One impression we got during the experiments was that some people only started to notice cybersickness effects in the moment they took off the HMD and the minutes after.
This might open up a new dimension as this cannot be countered by the proposed method, at least not with subjective measures, but maybe with biophysiological ones.

\subsection{Individual Factors}
In the presented experiment, there was no (initial) calibration or influence based on demographic data.
However, we analyzed the individual factors and made some finds. 
In contrast to existing research \cite{kim2005}, we found no correlation between the subjective vulnerability to motion sickness and measured cybersickness.
However, we based this only on one simple question asking if people suffer from motion sickness (given an explanation what this is) with the possibility of answering, \textit{no}, \textit{yes but rarely} and \textit{yes, often}.
We did the same for cybersickness with one additional option to answer: \textit{I never used a VR system}, and found a strong correlation to the sickness measures.
Our observation suggests two things: First, with the rise of consumer VR, people often already had contact to VR hardware, at least within the age group we studied and on a university campus, and second, they know whether they are vulnerable to cybersickness or not.
Finally, we found no correlation between age and sickness, and no effect of VR experience and sickness, different to existing research \cite{Freitag2016, Arns2005}, which both might be caused by the distribution of the study participants.

\subsection{Limitations}
In addition to the limitations of the experiment that were already mentioned, there are two more that serve as a foundation to future work.
First, the FoV reduction was performed evenly on the horizontal and on the vertical axis, which might have different effects.
Second, it should be more beneficial to already start with a slightly restricted FoV, as it can quickly be opened up, if the user feels good and has a faster effect if not.

\section{Conclusion\label{sec::conclusion}}
In this work, we evaluated the impact of sickness induced field of view reduction on cybersickness and presence measures.
We found no significant effects on cybersickness, nor an effect on presence.
Nevertheless, the experiment revealed some promising options for further investigations.
The most promising starting point seems to be finding a possibility to react faster to upcoming sickness and searching for a more reliable tracking of sickness. 
In any case, this topic gets more relevant with the increasing FoV of consumer HMDs and the rising interest in the serious and productive application of VR.

\acknowledgments
{
The authors would like to acknowledge the support by the Excellence Initiative of the German federal and state governments, the J\"ulich Aachen Research Alliance -- High-Performance Computing and the Helmholtz portfolio theme ``Supercomputing and Modeling for the Human Brain''. This project has received funding from the European Union's Horizon 2020 research and innovation programme under grant agreement No 720270 (HBP SGA1).
}

\bibliographystyle{abbrv-doi}

\bibliography{bib}
\end{document}